\documentstyle[12pt]{article}

\begin{document}
\title{Pattern formation during Rayleigh-B\'enard convection in
non-Boussinesq fluids}
\author{Hao-wen Xi, J.D. Gunton\\
Department of Physics\\
Lehigh University\\
Bethlehem, Pennsylvania 18015,\\
\\
and\\
\\
Jorge Vi\~nals\\
Supercomputer Computations Research
Institute, B-186\\
Florida State University\\
Tallahassee, Florida 32306-4052\\
and\\
Department of Chemical Engineering, B-203\\
FAMU/FSU College of Engineering\\
Tallahassee, Florida 32316-2175.}
\maketitle
\newpage

\section*{Abstract}

Motivated by recent experimental studies of Bodenschatz et al. [E.
Bodenschatz, J.R. de Bruyn, G. Ahlers and D.S. Cannell, Phys. Rev. Lett.
{\bf 67}, 3078 (1991) ], we present a numerical study of a
generalized two dimensional Swift-Hohenberg equation to
model pattern formation in
Rayleigh-B\'enard convection in a non-Boussinesq fluid.
It is shown that many of the features observed in these experiments can
be reproduced by this generalized model
that explicitly includes non-Boussinesq and mean flow effects.
The spontaneous formation of hexagons,
rolls, and a rotating spiral pattern is studied,
as well as the transitions and
competition among them. Mean flow, non-Boussinesq effects,
the geometric shape of the lateral wall, and sidewall
forcing are all shown to be crucial in the formation of the
rotating spirals. We also study nucleation and growth of hexagonal patterns
and find that the front velocity in this two dimensional model is consistent
with the prediction of marginal stability theory for one dimensional fronts.

\newpage

\section{Introduction}

One of the most striking examples of spatio-temporal self-organized
phenomena in nonequilibrium systems is
the rotating spiral states observed
in chemical and biological systems \cite{re:cb91}.
It is remarkable
that such time dependent but macroscopically coherent states can be
sustained in systems that are not in equilibrium.
The Belousov-Zhabotinsky reaction \cite{re:bz91}, for example,
has received
considerable attention as an example of a chemical wave propagation.
Spiral patterns in this system result from the
coupling of reaction and transport processes.

Recently, similar rotating spiral patterns have been
observed in Rayleigh-B\'enard convection in non-Boussinesq
fluids in large aspect ratio systems \cite{re:bo91}.
According to the classical work of Busse \cite{re:bu67},
the first bifurcation from the conducting state is to a convective state
of hexagonal symmetry. Convective cells form a stationary
honeycomb structure. Further away from threshold, the system undergoes
a new bifurcation to a state comprising parallel convective rolls (roll
patterns are the only patterns predicted and observed within the
Boussinesq approximation. The existence of of a stationary pattern of
hexagonal symmetry is a direct consequence of deviations from the
Boussinesq approximation). The predicted bifurcation
from hexagons to rolls is direct, so that
the fluid is expected to evolve to a stationary patterns of rolls.
Recently,
however, experiments on convection in CO$_{2}$ gas by Bodenschatz et
al. \cite{re:bo91} show that the system has a
tendency to spontaneously
form rotating spirals. These rotating states are long lived, and do
not decay
to the expected pattern of concentric rings (in a circular geometry).
Furthermore, depending on the value of the Rayleigh number, spirals
with a different number of arms have been observed.

Rayleigh-B\'enard convection in monocomponent fluids is
governed by the full three dimensional
fluid equations.
Because of the difficulty in solving the three dimensional
initial value problem posed by
the fluid equations, we and others have focused on the study of
simpler two dimensional model equations.
An example of such models is the so-called \lq\lq Swift-Hohenberg" (SH)
equation,
which is asymptotically equivalent to the long distance and long time
behavior of the fluid equations near onset of convection and in the
Boussinesq approximation \cite{re:sw77,re:co90,re:ma90}.
A great deal of theoretical and numerical
work on this latter model has been done
by Cross \cite{re:cr80}, and by Greenside et al. \cite{re:gr82},
but much remains to be
done in non-Boussinesq systems, even within the framework
of a SH-type equation.

We study in this paper a generalized SH equation
that includes both a quadratic nonlinearity and coupling to mean flow
effects.
The values of the parameters that enter the equation have been chosen to
be in the range appropriate for the experiments of Bodenschatz et al. We
find that stable rotating spirals are spontaneously formed during the
hexagon to roll transition, in agreement with the experimental
observations. The quadratic nonlinearity in the equation is
responsible for the rotational symmetry breaking and leads, by itself,
to stationary spiral patterns. When coupling to mean flow in included,
and therefore when the model equation is not potential, the same
transition leads to rotating spirals instead. However, spirals are not
obtained if one sets the quadratic term equal to zero but keeps the mean
flow term. Sidewall forcing also seems to be essential in obtaining this
pattern. Otherwise, rolls that are locally perpendicular to the sidewall
appear and no uniformly rotating state is observed.
Finally, once the spiral state is formed, it is
unstable to the removal of the quadratic nonlinearity in the equation.
The spiral state quickly decays to a set of concentric rings.

We note that Bestehorn et al. \cite{re:be92} have reported a numerical
study of this very same model to study the rotation of a spiral pattern,
but their study is limited to the special case in which the
initial configuration is a spiral. The fundamental
question addressed in our work is, in contrast, how the spiral pattern
is spontaneously formed during the hexagon to roll transition.

In Section II, we give a brief description of the theoretical model
used to describe pattern formation in non-Boussinesq systems,
and in Section III we provide a detailed description of the numerical
method for solving the model equation. In Section IV, we present various
numerical results, show in detail the
hexagonal and spiral patterns and compare our results with the experiments
of Bodenschatz et al. In Section V, we present a brief summary.

\section{A two dimensional model equation for convection in
non-Boussinesq fluids}

A great deal of our current understanding of the spatial and temporal
properties
of convective patterns near onset has been obtained by using model equations
such as the Swift-Hohenberg (SH) equation, which
in dimensionless variables takes the following
simple form \cite{re:sw77,re:ah81},
\begin{equation}
\label{eq:sh}
\frac{\partial \psi (\vec{r},t)}{\partial t} =
\left[ \epsilon -
\left( \nabla^{2} + 1\right)^{2} \right] \psi - \psi^{3},
\end{equation}
where $\vec{r}$ is a two dimensional vector lying in the $(x,y)$ plane
(the convective cell is parallel to this plane), and
$\epsilon$ is the control parameter.
This equation describes the evolution of
a single scalar field $\psi$ which is commensurate with the convective
rolls. When $\epsilon > 0$, this equation has roll-like solutions with
a wavenumber $q=1$. This equation can
also be written in potential form,
\begin{equation}
\frac{\partial \psi}{\partial t}=- \frac{\delta {\cal L}}{\delta \psi},
\end{equation}
where ${\cal L}$ is a Lyapunov functional. The dynamical
evolution of $\psi$ is naturally interpreted in terms of the
minimization of this functional.
Therefore a potential system cannot display either the oscillatory or
aperiodic time dependence observed in the experiments.
Although potential systems are quite
important for the understanding of the features of the emerging patterns,
an interesting question one may ask is which flow component is necessary
to describe the observed non-potential behavior. One such component
is the so-called large scale mean flow,
which was first proposed by Siggia and Zippelius \cite{re:si81}.

Large scale
mean flows are composed of one particular spatial harmonic of the basic roll
pattern. In the case of free-free boundary conditions, this harmonic
corresponds to a flow that is independent of $z$,
the coordinate normal to the convective cell plates.
This flow is not damped in the (unrealistic) case of free-free boundary
conditions, and is only slightly damped in the more realistic case of
rigid-rigid boundary conditions.
Since the nonlinear term in the fluid equations that gives rise to
large scale mean flow is not $\vec{v} \cdot \nabla \theta$, but
$\vec{v} \cdot \nabla \vec{v}$, the magnitude of the large scale
flow is inversely proportional to the
Prandtl number. Moreover, for a perfect straight roll
pattern, no large scale mean flow is generated; it appears
only when rolls are bent or modulated. In that case, the characteristic
length scale of these flows is large compared with the roll
wavelength. For example, the large scale mean flow contribution to
the SH equation \cite{re:si81,re:ma83} has been shown to play
a key role in the onset of weak turbulence in Boussinesq
systems \cite{re:gr88}.

We next describe a two dimensional model of convection
in a non-Boussinesq fluid. We use the
two dimensional generalized Swift-Hohenberg (GSH)
equation \cite{re:sw77,re:gr82,re:ma83},
given by Eqs. (\ref{eq:nonsh}) and
(\ref{eq:mean}) below, which we solve
by numerical integration. Our model in dimensionless units is defined by,
\begin{equation}
\label{eq:nonsh}
\frac{\partial \psi (\vec{r},t)}{\partial t} +
g_{m} \vec{U} \cdot \nabla \psi=
\left[ \epsilon - \left( \nabla^{2} + 1 \right)^{2} \right] \psi
-g_{2} \psi^{2} - \psi^{3} + f(\vec{r}),
\end{equation}
\begin{equation}
\label{eq:mean}
\left[ \frac{\partial }{\partial t} -Pr( \nabla^{2}-c^{2} ) \right]
\nabla^{2} \xi = \left[ \nabla(\nabla^{2} \psi \times
\nabla \psi ) \right]
\cdot \hat{e}_{z},
\end{equation}
where,
\begin{equation}
\vec{U}=(\partial_{y} \xi) \hat{e}_{x} - (\partial_{x} \xi) \hat{e}_{y},
\end{equation}
with boundary conditions,
\begin{equation}
\label{eq:bcs}
\xi|_{B}=\hat{n} \cdot \nabla \xi |_{B} = \psi|_{B}
= \hat{n} \cdot \nabla \psi |_{B} = 0,
\end{equation}
where $\hat{n}$ is the unit normal to the boundary of the domain of
integration, $B$. If the coupling to large scale
flow is dropped ($g_{m}=0$), Eq. (\ref{eq:nonsh}) is potential.
Also this equation reduces to the dimensionless SH
equation (Eq. (\ref{eq:sh})) when the coupling coefficients
$g_{2}=g_{m}=0$. As is the case in Eq. (\ref{eq:nonsh}), the scalar field
$\psi(\vec{r},t)$ is proportional to a linear combination of
the fluid temperature modulation
and vertical velocity fields at a point $\vec{r}$ in the
midplane of the fluid layer, parallel to the upper and lower walls of the
convective cell. $\xi(\vec{r},t)$ is the vertical vorticity
potential \cite{re:si81,re:ma83,re:gr88}.
The quantity $\epsilon$ is the control parameter.
$Pr$ is the Prandtl number of the fluid,
$\hat{e}_{x}$ and $\hat{e}_{y}$ are two unit vectors parallel to the
$x$ and $y$ direction respectively, and $c^{2}$ is
an unknown constant. A phenomenological forcing
field $f$ has been
included in Eq. (\ref{eq:nonsh}) to simulate lateral sidewall forcing
produced by
horizontal temperature gradients present in the experiments. As in earlier
studies \cite{re:xi91,re:vi91}, we have varied the strength and spatial
extent of $f$ in order to match the experimental observations.
In order to estimate the values of the various dimensionless
parameters that enter the GSH equation
in terms of experimentally measurable quantities,
we have derived a three mode amplitude equation
(see the Appendix for details). From the experiments described in reference
\cite{re:bo91}, we estimate that $g_{2}=0.35$, $g_{m}=50$,
$c^{2}=10$ and $Pr = 1$.
 The value of $\epsilon$ used in the
numerical calculation is related to the experimental value $\epsilon_{expt}$
by $\epsilon_{expt}=0.3594 \epsilon$.

\section{Numerical method}

The numerical method for solving the GSH equation (Eqs. (\ref{eq:nonsh})
- (\ref{eq:bcs})) is based on the elegant work
by Greenside et al. \cite{re:gr82}, and  Bj\o rstad et al. \cite{re:bj84}.
We sketch below
their numerical scheme. The key step is to recognize that the GSH equation
can be solved by the repeated solution of the
$\it linear$ constant coefficient biharmonic equation for a function $u$,
\begin{equation}
\label{eq:biharmonic}
(\nabla^{2}\nabla^{2}+a \nabla^{2}+b)u=f_{1},
\end{equation}
with boundary conditions,
\begin{equation}
u|_{R} = f_{2},   \ \  \hat{n} \cdot \nabla u|_{R} = f_{3},
\end{equation}
\noindent on a circular domain $R$. The constants $a$ and $b$ are real numbers,
$\hat{n} \cdot \nabla $ denotes the normal derivative taken at the
boundary of the domain, and
$f_{1}$ , $f_{2}$ and $f_{3}$ are given functions.
For the case of rigid boundaries, we have $f_{2}=f_{3}=0$.

Consider an $\it implicit$ backward Euler
discretization scheme in time, to yield the following finite difference
set of equations for Eqs. (\ref{eq:nonsh})-(\ref{eq:bcs}),
$$
\frac{\psi(\tau+\Delta \tau)-\psi(\tau)}{\Delta \tau}
+g_{m} \vec{U}(\tau+\Delta \tau) \cdot \nabla \psi(\tau+\Delta \tau) =
$$
\begin{equation}
\label{eq:disc1}
L\psi(\tau+\Delta \tau)
-g_{2}\psi^{2}(\tau+\Delta \tau)-\psi^{3}(\tau+\Delta \tau),
\end{equation}
$$
\frac{\nabla^{2} \left[ \xi(\tau+\Delta \tau)-\xi(\tau) \right] }{\Delta \tau}
+Pr \nabla^{2} \left[ c^{2} - \nabla^{2} \right]
\xi(\tau+\Delta \tau)=
$$
\begin{equation}
\label{eq:disc2}
-\left[ \nabla \psi(\tau+\Delta \tau) \times
\nabla(\nabla^{2} \psi(\tau+\Delta \tau) \right]
\cdot \hat{e}_{z},
\end{equation}
where $\psi(\tau)$ and $\xi(\tau)$ are the known solutions of
Eqs. (\ref{eq:nonsh})-(\ref{eq:mean}) at time $\tau$, $\Delta \tau$ is the
time step,
$ L=\epsilon-(1+\nabla^{2})^{2} $ is a linear biharmonic
operator, and $\psi(\tau+\Delta \tau)$ and $\xi(\tau+\Delta \tau)$ are
the unknown implicit solutions at the next time step ($\tau+\Delta \tau$)
(we have temporarily suppressed the spatial arguments).
We solve Eqs. (\ref{eq:disc1}) and (\ref{eq:disc2}) by using a
multi-iteration Gauss-Seidel scheme. We
first assume that $\psi(\tau +\Delta \tau)$
and $\xi(\tau+\Delta \tau)$
are obtained by successive approximations of the form,
\begin{equation}
\label{eq:approx1}
\psi(\tau+\Delta \tau) \simeq \psi_{k+1}=\psi_{k}+\delta_{k}, ~~~~
{\rm with} \psi_{0}=\psi(\tau),
\end{equation}
\begin{equation}
\label{eq:approx2}
\xi(\tau+\Delta \tau) \simeq \xi_{k+1}=\xi_{k}+\theta_{k}, ~~~~
{\rm with} \xi_{0}=\xi(\tau),
\end{equation}
where $\psi_{k}$ and $\xi_{k}$ are the approximations
at the $k-th$  iteration. The equation satisfied by
the so-called outer correction fields, $\delta_{k}$
and $\theta_{k}$, can be obtained (assuming
$\parallel \delta_{k} \parallel$  $\ll$  $\parallel \psi_{k} \parallel$
and  $\parallel \theta_{k} \parallel$  $\ll$  $\parallel \xi_{k} \parallel$
in the maximum  norm). By substituting Eqs. (\ref{eq:approx1}) and
(\ref{eq:approx2}) into Eqs. (\ref{eq:disc1}) and (\ref{eq:disc2}),
and linearizing them with respect to
 $\delta_{k}$ and $\theta_{k}$, we then obtain the
standard Gauss-Seidel
iteration scheme for the unknown corrections $\delta_{k}$ and $\theta_{k}$,
\begin{equation}
\label{eq:corr1}
\left[ L-3\psi_{k}^{2}-2g_{2}\psi_{k}-\frac{1}{\Delta \tau} \right]\delta_{k}=
\frac{\psi_{k}-\psi_{0}(\tau)}{\Delta \tau}
+g_{m} \vec{U}_{k} \cdot \nabla \psi_{k} -(L\psi_{k}-\psi_{k}^{3}
-g_{2}\psi_{k}^{2}),
\end{equation}
$$
\nabla^{2} \left[ \nabla^{2}
-h^{2} \right] \theta_{k}
=-\frac{1}{Pr \Delta \tau} \nabla^{2}\xi_{0}
- \nabla^{2} \left[ \nabla^{2} -h^{2}\right] \xi_{k}
$$
\begin{equation}
\label{eq:corr2}
+\frac{1}{Pr}\left[ \nabla \psi_{k+1} \times
\nabla(\nabla^{2} \psi_{k+1}) \right] \cdot \hat{e}_{z},
\end{equation}
with $h^{2}=(c^{2}+\frac{1}{Pr \Delta \tau})$.
The right hand sides of Eqs. (\ref{eq:corr1}) and (\ref{eq:corr2})
are the $k-th$ outer
residuals, $r_{outer}^{\psi}(k)$  and $r_{outer}^{\xi}(k)$ which measure
the extent to which Eqs. (\ref{eq:disc1}) and (\ref{eq:disc2}) are satisfied
by the $k-th$ order approximation.
Given the residuals, we solve for the outer corrections, $\delta_{k}$
and $\theta_{k}$, and then
obtain a better approximation to $\psi(\tau+ \Delta \tau)$
and $\xi(\tau+ \Delta \tau)$. Iteration continues over
the index $\it k$,
until both the outer residuals and the outer corrections are small compared
to $\psi_{k}$ and $\xi_{k}$, that is,
\begin{equation}
{\rm max} (\parallel \delta_{k} \parallel, \parallel r_{outer}^{\psi}(k)
\parallel)
\leq \epsilon_{rel} (\parallel \psi(k) \parallel) + \epsilon_{abs},
\end{equation}
and,
\begin{equation}
{\rm max}(\parallel \theta_{k} \parallel, \parallel r_{outer}^{\xi}(k)
\parallel)
\leq \epsilon_{rel} (\parallel \xi(k) \parallel) + \epsilon_{abs},
\end{equation}
where $\epsilon_{rel}$ and $\epsilon_{abs}$ are the relative
and absolute error tolerances, chosen to be 0.1 and $10^{-4}$ respectively
in our calculations. When the convergence criteria are satisfied,
$\psi_{k+1}$ and $\xi_{k+1}$ are set to be
$\psi(\tau+\Delta \tau)$ and $\xi(\tau+\Delta \tau)$. In the numerical
solution, we have found that $\Delta \tau > \Delta \tau_{max} \simeq 1.8 $
will cause a numerical instability, so we have chosen
$ \Delta \tau $ $<$ $\Delta \tau_{max}$.
($\Delta \tau \ll 1.0$ during the initial transient at onset).
For a given $\psi_{k+1}$, Eq. (\ref{eq:corr2}) has the exact form of Eq.
(\ref{eq:biharmonic}), with
$a=-c^{2}-\frac{1}{Pr \Delta \tau}$, $b=0$, $f_{1}=r_{outer}^{\xi}(k)$,
$f_{2}=f_{3}=0$, and can be solved rapidly
and accurately. Eq. (14) is almost of the form of Eq.
(\ref{eq:biharmonic}) except
for the non-constant term $-2g_{2} \psi_{k}-3\psi_{k}^{2}$ in the
left hand side operator. We can reduce this equation
to the desired constant coefficient
biharmonic form by assuming a successive approximation of the form:
\begin{equation}
\label{eq:fudge}
\delta_{k} \simeq \delta_{k,m+1}=\delta_{k,m}+\eta_{m},
\end{equation}
where the inner correction field, $\eta_{m}(x,y)$,
is assumed to be small compared to $\delta_{k,m}$, the $m-th$ approximation
to $\delta_{k}$. By substituting Eq. (\ref{eq:fudge}) into Eq.
(\ref{eq:corr1}) and solving for $\eta_{m}$ by
approximating the non-constant term acting on $\eta_{m}$ as a constant C,
we obtain
\begin{equation}
\label{eq:fudge2}
\left[ L-\frac{1}{\Delta \tau}+C \right] \eta_{m}=r_{outer}^{\psi}(k)
-\left[ L-\frac{1}{\Delta \tau}-2g_{2} \psi_{k}-3 \psi_{k}^{2} \right]
\eta_{k,m}.
\end{equation}
The right hand side is the $m-th$ inner residual,
$r_{inner}^{\psi}(k,m)$ of Eq. (\ref{eq:fudge2}).
This measures the extent to which
Eq. (\ref{eq:corr1}) is satisfied after $\it m$ iterations. The constant
$C = -2g_{2}<\psi_{k}>-3<\psi_{k}^{2}>$, where $<>$ denotes a spatial
average over the entire system.
Now Eq. (\ref{eq:fudge2}) has the exact form of the constant coefficient
biharmonic equation given in Eq.(\ref{eq:biharmonic}),
with $a=2$, $b=1+\frac{1}{\Delta \tau}-\epsilon-C$,
$f_{1}=-r_{inner}^{\psi}(k,m)$, and $f_{2}=f_{3}=0$.
The criterion for inner accuracy after $\it m$ iterations has the
following form \cite{re:gr82} \cite{re:bj84}:
\begin{equation}
\frac{ \parallel r_{inner}^{\psi}(k,m) \parallel }
{ \parallel r_{outer}^{\psi}(k) \parallel } \leq
\alpha \left[ \frac{ \parallel r_{outer}^{\psi}(k) \parallel }
{ \parallel r_{outer}^{\psi}(1) \parallel } \right]^{\beta} ,
\end{equation}
where $\alpha$ and $\beta$ are chosen to be 0.1 and 0.5 in
our numerical simulations, since the rate of convergence is not
sensitive to $\alpha$ and $\beta$ \cite{re:gr82,re:bj84}.

We describe next the discretization of the spatial derivatives
in Eqs. (\ref{eq:disc1})-(\ref{eq:disc2})
for the geometry of interest.
Since both the Laplacian and biharmonic operators have a
singularity at the origin of a polar coordinate system, we have
found it convenient
to use a Cartesian coordinate system and approximate the boundary conditions.
We have used the usual 5- and 13- point discretizations of the Laplacian
and biharmonic operators on a square grid of
$N \times N $ points, which is second-order accurate in the mesh spacing.
We approximate the circular boundary conditions on $\psi$ and $\xi$ by
taking $\psi (\vec{r},t) = \xi (\vec{r},t) = 0$
for $\| \vec{r} \| \ge D/2$, where $\vec{r}$ is the location of a node
with respect to the center of the domain of integration, and
$D$ is the diameter of the circular domain. This
approximation is presumably not dangerous since
boundary effects are known to affect the bulk behavior over length
scales of the order of
$\epsilon^{-\frac {1}{2}}$ \cite{re:ma83},
which is large compared to the spatial
discretization.
 Since the convective pattern has  locally periodic
solutions with wavelength $2\pi$, except near the boundary,
8 or 12 grid points per wavelength are used in our
numerical calculations.

Two different kinds of initial conditions are used.
The first one is a gaussian distributed random initial condition with
zero mean value and small variance (see below) for $\psi$,
and  $\xi$=0. This initial condition models
the conduction state. The second type of
initial conditions used is a solution
from a previous run ($\psi(t_{0})$, $\xi(t_{0})$).  This, for example,
would correspond to studying the
transition from a hexagonal state to a roll state in an
experiment that suddenly increased the Rayleigh number.

\section{Numerical results}

In this section, we report the results of our numerical
calculations based on Eqs. (\ref{eq:nonsh})-(\ref{eq:bcs}) in a
large aspect ratio cell, and compare the
pattern evolution obtained with the experiments of Bodenschatz et al. We begin
with the formation of a convecting state of  hexagonal
symmetry from the conducting
state. Next, we present numerical evidence for the
spontaneous formation of a rotating spiral pattern during the
hexagon to roll transition. We also report
numerical results on the formation of rotating spiral patterns during the
transition from conduction to rolls.

\subsection{Nucleation of a pattern with hexagonal symmetry}

We consider as initial condition $\psi( \vec{r},t=0)$ a Gaussian random
variable with zero mean and variance $10^{-6}$.
The forcing field chosen is $f(\vec{r})=0$, simply because there is no
influence from the lateral boundaries before the nucleated pattern
reaches the boundary. We also neglect in this case the mean flow field.
We numerically solve Eq. (\ref{eq:nonsh})
in a square domain of side $L = 128 \pi $ (in our dimensionless
variables, this corresponds to an aspect ratio $\Gamma = L/ \pi = 128$).
The differential equation is discretized on a square grid of
$512 \times 512$ nodes. We use $g_{2}=0.35$ and $\Delta x$=
$\pi /4.25$.  We take $\epsilon = 0.01$
except in a small square region near the center of the cell (of size $16
\times 16$ nodes) where $\epsilon = 0.055$.
This space dependent $\epsilon$ models a small localized inhomogeneity
in one of the cell plates.
According to the calculation by Busse \cite{re:bu67},
and our estimate of the values of the parameters in the GSH equation
(see the Appendix), a hexagonal pattern should be stable
for both $\epsilon=0.01$ and $\epsilon=0.05$. The temporal evolution of the
pattern is shown in Fig. 1. It presents an early transient
behavior during which a local convective region with hexagonal symmetry
has just nucleated.  Six fronts of rolls, traveling away from the hexagonal
patch located at the center, propagate into the conduction region.
This is qualitatively similar to the experimental observations of
Bodenschatz et al. \cite{re:bo91,re:bo92}.
The observation of nucleation and growth is
especially interesting since it provides an example of
competition among different symmetries, i.e., a uniform conduction state
as the background state, a region of hexagonal symmetry being nucleated and
rolls in the front region separating the two. This situation is also
interesting from the point of view of pattern selection during front
propagation in dimensions higher than one.
It is worth pointing out, however, that the shape of the
front region obtained numerically is
somewhat different than the experimental one.
This difference may be attributable to the fact that the numerical
value $\epsilon (\vec {r})$ used here is much larger than the
 experimental value. Unfortunately,
it is very time
consuming for us to solve the equation for the experimental
value $\epsilon_{exp} \sim 10^{-4}$.

We have also estimated the speed of
propagation of the front that separates the hexagonal pattern and the
uniform state. This speed, at the center of the planar sides, and
along their normal direction, is approximately constant in time and equals
$v_{\perp} = 0.37$.
The value given by marginal stability theory for the one dimensional
Swift-Hohenberg equation ($g_{2}=0$) is $v_{MS} = 0.397$
\cite{re:de83,re:sa88}. It will be interesting to measure the front velocity
in the experiment.

\subsection{Conduction to hexagon transition}

In what follows, we consider a circular cell of radius $r = 32 \pi$, which
corresponds to an aspect ratio of $\Gamma =r / \pi = 32$. A grid with
$N^{2}$ nodes has been used with spacing $\Delta x = \Delta y = 64
\pi$/N, and $N=256$, with an approximate boundary conditions
to mimic a circular cell as we discussed in section 3.
The initial condition $\psi (\vec{r},t=0)$ is again a random
variable, gaussianly distributed with zero mean and variance $10^{-1}$.
In this case $\epsilon = 0.1$, which is also within the region in
which a hexagonal pattern is stable. The forcing field $f=0$ everywhere
except at the nodes
adjacent to the boundary where $f = 0.1$.
This value is of the same order as a previous estimate obtained for a
similar experimental setup,
such that the convective heat current measured during a ramping
experiment agreed with the numerical solution of the SH equation
\cite{re:xi91,re:vi91}. Figure 2 presents
the evolution of the hexagonal pattern evolution from random
initial condition. Mean flow field effects have also been included in
the calculation.

\subsection{Hexagon to roll transition, and formation of rotating spirals}

We have used the configuration shown in Fig. 2(d)
as the initial condition, with exactly the same parameters and
forcing field
$f$ as before ($f=0.1$), but have increased $\epsilon$ very slowly up to
$\epsilon = 0.3$: $\epsilon=0.1+1.67\times 10^{-4}t$ for $0<t<1200$,
and $\epsilon=0.3$
for $t>1200$. Figure 3 shows a sequence of configurations
during the early transient regime of the hexagon to roll
transition.
 Rolls appear in the vicinity of the sidewall and tend to
orient themselves parallel to it. Defects glide toward each
other and invade nearby regions of hexagonal symmetry to create a region
of rolls that spreads across the cell as the transition proceeds.
The formation of spirals is already noticeable in Fig. 3(d).

Fig. 4 shows a continuation of the sequence of configurations shown in
Fig. 3. Fig. 4(a) shows how rolls bend
rapidly to form a locally disordered texture near the upper left portion
of the system.
Defects in the disordered texture
migrate away or annihilate each other, leaving a roughly
 uniform patch of rolls, and eventually annihilating themselves, ending in a
three-armed spiral (Figs. 4(c)-(e)).
The final state of a rotating spiral (Figs. 4(e) and (f)) is
remarkably similar to the one observed in the experiments, and occurs at
$t \simeq 49000 \simeq 12$ horizontal diffusion times \cite{re:note92}.
The corresponding experimental times are in the range of 10 to 20
horizontal diffusion times.

\subsection{Roll to hexagon transition}

Figures 5(a)-(d) show the evolution
from a
initial rotating spiral
to a hexagonal pattern. We have used the configuration shown in
Fig. 4(f) as the initial condition, have kept the same
parameters and forcing field as before ($g_{2}=0.35$, $g_{m}=50.0$,
$c^{2}=10.0$ and Pr=1.0), but have set $\epsilon = 0.1$.
Initially, regions of hexagonal symmetry
emerge near the tip and the end of the spiral, forming  local domains
of hexagons and then spreading across the cell. An interesting
feature worth noticing during this transient period is the relative
orientation between domains with different symmetries. If one uses the
outer layer of hexagons in the domain to define a boundary, one can see from
Fig. 5 that
there is a tendency for these boundaries to be perpendicular to the
direction defined by the rolls. In systems
in which the evolution is
governed solely by a Lyapunov functional,
theoretical arguments have been given to explain analogous orientation
phenomena \cite{re:cr80}. Although these arguments do not apply in the
present case, the patterns observed both in the experiments
and in our numerical calculations still show that, locally, the boundary
separating regions of hexagons and rolls tends to be perpendicular
to the rolls. Perhaps a coupled
Newell-Whitehead-Segel model (which has an associated Lyapunov functional)
could be used to describe the formation of
domain boundaries between hexagons and rolls \cite{re:bam90}.

\subsection{Conduction to roll transition}

We study in this section the formation of a set of convective roll
directly from the conducting state. We use a
random initial condition
(gaussianly distributed with zero mean and a variance of $10^{-4}$),
and set $\epsilon$=0.3.
Figure 6 shows that concentric rolls are created near the
sidewall and propagate inward. Small and large length
scale defects anneal out rapidly leaving a disordered structure at the center
of the cell (Fig. 6(a) and (b)). Further evolution involves the
annihilation of defects at the center of cell.
Figures 6(c) and (d) show a two-armed rotating spiral in which all
defects have been eliminated from the center of the cell. This calculation
shows the importance of sidewall forcing and the geometric shape of
the container for the formation of a rotating spiral.

Another interesting feature observed in the experiments that
our model can also reproduce is that a stable, $n$-armed spiral tends toward
one with fewer arms when $\epsilon$ is decreased.
With the same random initial condition and parameters as in
Fig. 6, we have obtained a two-armed spiral for $\epsilon=0.3$
(Fig. 6), a one-armed spiral for $\epsilon=0.26$ (Fig. 7), and
a zero-armed spiral (concentric rolls) for $\epsilon=0.22$.

\subsection{Stability of the rotating spiral pattern}

We discuss in this section the condition under
which the rotating spiral is stable. Earlier
work \cite{re:xi92} established that a spiral pattern can be
spontaneously formed in the absence of mean flow
$(g_{m}=0)$, but retaining
the non-Boussinesq contribution $(g_{2}>0)$.
The spiral pattern obtained, however, is stationary.
The addition of mean flow effects is sufficient to spontaneously produce
a uniformly rotating spiral.
We address here the stability of an already formed rotating
spiral with respect to changes in the various terms of the GSH
equation.

Once the rotating spiral has been formed, we set $g_{2}=0.0$.
We  use the configuration shown in Fig. 7(b)
as the initial condition, with exactly the same parameters
($\epsilon=0.26$, $c^{2}=10.0$, $g_{m}=50.0$, $Pr=1.0$) and
forcing field $f$ as in Fig. 7(b).
Figure 8 shows the resulting evolution of the pattern.
The defect in the arms of the spiral propagates
inwards as the one-armed
spiral  evolves to a set of concentric rolls.
This result suggests that non-Boussinesq effects are needed
to stabilize the spiral structure,
even under the presence of mean flow. It would be very useful
to replicate this observation experimentally. Further, if $f=0$
rolls tend to align themselves normal to the sidewall,
and a pattern of almost straight and parallel rolls obtains.

In summary, sidewall forcing and non-Boussinesq effects ($g_{2}>0$)
are essential to produce a spiral pattern. A rotating
spiral is obtained when $g_{m} \neq 0$. Once the rotating
spiral is formed, if $g_{2}=0, g_{m} \neq 0$, the spiral decays to
concentric rolls.

\subsection{Convective current versus the number of arms in a spiral pattern}

We have observed that for a given value of $\epsilon$, the number
of arms of the rotating spiral depends on the initial
condition. In order to compare the
convective heat current for the same value of $\epsilon$,
but for spiral patterns containing different numbers of
arms, we proceed in the following
way: the configurations shown in Fig. 7a (zero armed spiral
for $\epsilon=0.22$),
Fig. 7b (one-armed spiral for $\epsilon=0.26$),
and Fig. 6d (two-armed spiral for $\epsilon=0.30$) are
taken as initial conditions and $\epsilon$ is set
equal to $\epsilon=0.26$. The three spiral pattern remain stable in all
cases. We then calculate the convective heat current
as a function of the number of arms in a spiral pattern.
In Fig. 9, we compare the spatial and temporal
average of the convective current $<J_{conv}>$
for the three cases discussed above. It is apparent from Fig. 9 that
$< J_{conv}>$ decreases with increasing the number of arms in
the spiral. Since the Nusselt number is related to the convective current by
Nu=$J_{conv}+1$, it would also be interesting to check
this observation experimentally.

\section{Conclusions}

We have investigated
a model of convection for non-Boussinesq fluids that allow
patterns of various symmetries.
The model used is a generalization of the Swift-Hohenberg equation that
includes a quadratic term and coupling to large scale mean
flows. The parameters in the equation have been
chosen to match the experiments of Bodenschatz et al. on CO$_{2}$ gas
\cite{re:bo91}.
An appropriate value for the control
parameter $\epsilon$ takes the conduction state to
an ordered hexagonal state analogous to the ones observed in experiments.
We then show that upon
increasing $\epsilon$, the
hexagonal state evolves into a
new roll state that contains a rotating spiral pattern.
The time scale to form a rotating spiral is on the
time scale of 10 horizontal diffusion time.
These results are also in good agreement with the experimental studies
on CO$_{2}$ gas.
The observation of the stationary rotating spiral pattern
is not in predicated by Busse \cite{re:bu67}.
According to Busse \cite{re:bu67},
when the control parameter exceeds the
bifurcation point of the hexagon-roll state,
the system is expected to evolve to a
stationary parallel roll
state.
This is because they studied the convective
fluid in an infinite cell. We have
given a preliminary study of the mechanism
for spiral pattern formation.
Our calculations  illustrate the
strong influence of non-Boussinesq effects, sidewall forcing,
and mean flow on the appearance and stability of the
rotating spiral patterns.
Our results for convective current of
different armed-spirals show that the final
state of the rotating spiral dependents on the
initial configuration.
We have seen that a zero-armed, a one-armed
and a two-armed  rotating spiral
pattern could exist for the same control parameter.
This suggests
the existence of  multi-states for a given control parameter.
Our numerical results show that the
two dimensional generalized Swift-Hohenberg
quation can  describe
quantitatively detail the three dimensional
convective dynamics of a fluid beyond the Boussinesq approximation.

We conclude by suggesting some further analytical
and numerical studies. It would be very useful to study how
the instabilities, such as the
zigzag, cross-roll, and Eckhaus instabilities
are affected by the influence of the non-Boussinesq effect
and the finite size of the boundaries.
It would be also very interesting to study the
dynamics of the core and the tip (dislocation) of a rotating spiral, and to
determine the speed of climbing motion of the tip.

\section*{Acknowledgments}

We wish to thank E. Bodenschatz, G. Ahlers and D. Cannell for suggesting
the numerical investigation of the generalized Swift-Hohenberg equation,
and them and H.S. Greenside, P. Hohenberg for many stimulating
conversations and comments. This work was supported in part
by the National Science Foundation under Grant No. DMR-9100245.
This work is also supported in part by the Supercomputer Computations
Research Institute, which is partially funded by the U.S. Department of Energy
contract No. DE-FC05-85ER25000. The calculations reported
here have been carried out on the Cray Y-MP at
the Pittsburgh Supercomputing Center.

\section*{Appendix}

In this appendix a detailed analysis of the various stationary solutions
of the generalized Swift-Hohenberg equation is presented.
In dimensionless units \cite{re:ah81}, the generalized SH model
can be written as,
\begin{equation}
\label{eq:gsh1}
\tau_{0}\frac{\partial \psi (\vec{r},t)}{\partial t} =
\left[ \epsilon -  \frac{\xi_{0}^{2}}{4q_{c}^{2}} \left( \nabla^{2}
+ q_{c}^{2} \right)^{2} \right] \psi
 -g_{2} \psi^{2} -g_{3} \psi^{3}.
\end{equation}
We set \cite{re:ah81},
\begin{equation}
\psi= \sqrt{2} \sum_{j=1}^{3} Re(A_{j}) e^{i \theta_{j}} ,
\end{equation}
where $\theta_{j}= \vec{q}_{j} \cdot \vec{r}$,
$\sum_{j=1}^{3} \theta_{j}=0 $ and $A_{j}$ is a complex amplitude.

If we substitute $\psi$, $\psi^{2}$ and $\psi^{3}$ into
Eq.(\ref{eq:gsh1}) and keep
the lowest order in the amplitudes, we obtain
\begin{equation}
\label{eq-003}
\tau_{0}\frac{\partial A_{1}}{\partial t} =
\left[ \epsilon - \xi_{0}^{2} \left( \frac {\partial} {\partial x_{1}}
- \frac {i}{2q_{c}} \frac{\partial^{2}}{\partial y_{1}^{2}}  \right)^{2}
 \right] A_{1} - a A_{2}^{*}A_{3}^{*} - b A_{1}( |A_{2}|^{2}
+|A_{3}|^{2} ) - c A_{1} |A_{1}|^{2},
\end{equation}
\begin{equation}
\tau_{0}\frac{\partial A_{2}}{\partial t} =
\left[ \epsilon - \xi_{0}^{2} \left( \frac {\partial} {\partial x_{2}}
- \frac {i}{2q_{c}} \frac{\partial^{2}}{\partial y_{2}^{2}}  \right)^{2}
 \right] A_{2} - a A_{1}^{*}A_{3}^{*} - b A_{2}( |A_{1}|^{2}
+|A_{3}|^{2} ) - c A_{2} |A_{2}|^{2},
\end{equation}
\begin{equation}
\tau_{0}\frac{\partial A_{3}}{\partial t} =
\left[ \epsilon - \xi_{0}^{2} \left( \frac {\partial} {\partial x_{3}}
- \frac {i}{2q_{c}} \frac{\partial^{2}}{\partial y_{3}^{2}}  \right)^{2}
 \right] A_{3} - a A_{1}^{*}A_{2}^{*} - b A_{3}( |A_{1}|^{2}
+|A_{3}|^{2} ) - c A_{3} |A_{3}|^{2},
\end{equation}
where $a=\sqrt{2} g_{2}$, $b=3g_{3}$ and $c=\frac {3}{2} g_{3}$
and,
\begin{equation}
\frac {\partial}{\partial x_{j}}=cos\theta_{j} \frac {\partial}
{\partial x}+sin\theta_{j} \frac {\partial}{\partial y},
\end{equation}
\begin{equation}
\frac {\partial}{\partial y_{j}}=-sin\theta_{j} \frac {\partial}
{\partial x}+cos\theta_{j} \frac {\partial}{\partial y},
\end{equation}
\begin{equation}
\theta_{1}=0, \ \  \theta_{2}=\frac {2 \pi} {3}
\ \ and  \ \ \theta_{3}=\frac {4 \pi} {3}.
\end{equation}
If we assume a uniform solution, we simply have,
\begin{equation}
\label{eq:amp1}
\tau_{0}\frac{\partial A_{1}}{\partial t} =
\epsilon A_{1} - a A_{2}^{*}A_{3}^{*} - b A_{1}( |A_{2}|^{2}
+|A_{3}|^{2} ) - c A_{1}|A_{1}|^{2},
\end{equation}
\begin{equation}
\label{eq:amp2}
\tau_{0}\frac{\partial A_{2}}{\partial t} =
\epsilon A_{2} - a A_{1}^{*}A_{3}^{*} - b A_{2}(|A_{1}|^{2}
+|A_{3}|^{2} ) - c A_{2} |A_{2}|^{2},
\end{equation}
\begin{equation}
\label{eq:amp3}
\tau_{0}\frac{\partial A_{3}}{\partial t} =
\epsilon A_{3}- a A_{1}^{*}A_{2}^{*} - b A_{3}( |A_{1}|^{2}
+|A_{3}|^{2} ) - c A_{3} |A_{3}|^{2}.
\end{equation}
This set of equations can be written in  variational form as,
\begin{equation}
\tau_{0}\frac{\partial A_{i}}{\partial t}=- \frac{\delta
{\cal L} }{\delta A_{i}^{*}}.
\end{equation}
where the Lyapunov functional $\cal L$ is given by,
\begin{equation}
\begin{array}{l}
{\cal L}=-\epsilon(|A_{1}|^{2}+|A_{2}|^{2}+|A_{3}|^{2}))
+a(A_{1}^{*}A_{2}^{*}A_{3}^{*}+A_{1}A_{2}A_{3}) \\
+b(|A_{1}|^{2}|A_{2}|^{2}+|A_{2}|^{2}|A_{3}|^{2}+|A_{3}|^{2}|A_{1}|^{2})
+\frac{c}{2}(|A_{1}|^{4}+|A_{2}|^{4}+|A_{3}|^{4}).
\end{array}
\end{equation}

The dynamical system Eqs. (\ref{eq:amp1})-(\ref{eq:amp3})
has three stationary and homogeneous solutions:
\begin{equation}
\begin{array}{ll}
Conduction ~ state: & \\
 & A_{1}=A_{2}=A_{3}=0. \\
 & \\
Hexagonal ~ state: & \\
 & A_{1}=|A|e^{i\theta_{1}}, A_{2}=|A|e^{i\theta_{2}},
A_{3}=|A|e^{i\theta_{3}}, \\
 & \theta_{1}+\theta_{2}+\theta_{3}=0, \\
 & |A|=(-a-\sqrt{a^{2}+4(2b+c)\epsilon})/(2(2b+c)). \\
 & \\
Roll ~ state: & \\
 & A_{1}=|A|e^{i\theta_{1}}, \ \  A_{2}=A_{3}=0, \\
 & |A|=\sqrt{\epsilon/c}. \\
\end{array}
\end{equation}

The linear stability of these solutions is determined by the
eigenvalues of the
matrix $\delta^{2} {\cal L}/ \delta A_{i} \delta A_{j}$,
linearized around the stationary solutions.
For $\epsilon_{a} \leq \epsilon \leq 0$, both conduction and hexagons
are stable; for $0 \leq \epsilon \leq \epsilon_{r}$, only
hexagons are stable; for $\epsilon_{r} \leq \epsilon \leq \epsilon_{b}$
both hexagon and roll are stable; and for $\epsilon \geq \epsilon_{b}$,
only rolls are stable. From the corresponding values of the
stability boundaries obtained for the amplitude equation \cite{re:pe88},
we find,
\begin{equation}
\label{eq:ea}
\epsilon_{a}=-\frac{a^{2}}{(8b+4c)}=-\frac{2g_{2}^{2}}{15g_{3}},
\end{equation}
\begin{equation}
\label{eq:er}
\epsilon_{r}=\frac{a^{2}c}{(b-c)^{2}}=\frac{4g_{2}^{2}}{3g_{3}},
\end{equation}
\begin{equation}
\label{eq:eb}
\epsilon_{b}=\frac{a^{2}(b+2c)}{(b-c)^{2}}=\frac{16g_{2}^{2}}{3g_{3}}.
\end{equation}

Equations (\ref{eq:ea}-\ref{eq:eb}) can be compared
with the corresponding
equation obtained by the Busse \cite{re:bu67}.
This allows to determine the coefficients $a,b,c$
in the amplitude equation as
\begin{equation}
a^{2}=\frac{3P^{2}}{R_{c}},
\end{equation}
\begin{equation}
b=\frac{3\Re_{h}-\Re_{r}}{2},
\end{equation}
\begin{equation}
c=\Re_{r}
\end{equation}

Here $\Re_{h}$, $\Re_{r}$ and $P$ are the constants.
In the case a rigid-rigid boundary layer, we have
\begin{equation}
R_{c}=1707,
\end{equation}
\begin{equation}
\Re_{h}=0.8936+0.04959Pr^{-1}+0.06787Pr^{-2},
\end{equation}
\begin{equation}
\Re_{r}=0.69942-0.00472Pr^{-1}+0.00832Pr^{-2}.
\end{equation}
where $Pr$ is the Prandtl number, $P$ is the
non-Boussinesq parameter \cite{re:bu67}.

Since the system of equations (\ref{eq:amp1}-\ref{eq:amp3})
has an associated potential $\cal L$,
the absolute stable
state corresponds to the global minimum of $\cal L$, while metastable states
correspond to local minima. The existence of the Lyapunov functional
ensures that two stable phases can coexist only when they have the same
value of $\cal L$. We define $\epsilon_{T}$ to be the value for which
hexagons and the conduction state coexist,
and $\epsilon_{T'}$ the value fot the coexistence of hexagons and rolls.
We obtain,
\begin{equation}
\epsilon_{T}=-\frac{8}{9} \epsilon_{a} ,
\end{equation}
\begin{equation}
\epsilon_{T'}=(2b+c)|A|^{2}+a|A|,
\end{equation}
where $|A|$ is the solution of
\begin{equation}
3c(2b+c)|A|^{3}+2ac|A|^{2}-(2b+c)|A|-a=0.
\end{equation}

In order to obtain the values of the coupling coefficients $g_{2}$ and
$g_{3}$, we calculate the convective current
$J_{conv}$ for the various patterns. $J_{conv}$
is given by \cite{re:ah81},
\begin{equation}
\label{eq-023}
J_{conv}=\sum_{i} |A_{i}|^{2}.
\end{equation}
For a hexagonal pattern $J_{conv}$ is,
\begin{equation}
\begin{array}{l}
\label{eq-024}
J_{conv}=3|A_{hex}|^{2}=
3 \left[ \left( a+ \sqrt{a^{2}+4(2b+c)\epsilon} \right) / (4b+2c) \right]^{2}\\
        \\
=3\left[
(\sqrt{2}(g_{2}^{2}/\sqrt{g_{3}})^{1/2}+\sqrt{2g_{2}^{2}/g_{3}+30\epsilon})
/(15 \sqrt{g_{3}}) \right]^{2}.
\end{array}
\end{equation}
For a roll pattern $J_{conv}$ is
\begin{equation}
\label{eq-025}
J_{conv}=|A_{roll}|^{2}=\frac{\epsilon}{c}=\frac{2\epsilon}{3g_{3}}.
\end{equation}

By substituting the experimental values \cite{re:bo91}
of the threshold for $\epsilon_{a} \approx -2.3 \times 10^{-3}$
in the previous expressions for the stability boundaries of the various
patterns, we obtain, $g_{2}^{2} /g_{3} \simeq 0.0345$.
Furthermore, at $\epsilon=0.02$, we have (from Fig. 2 in \cite{re:bo91}),
that $\epsilon_{r} \simeq 0.06$; hence
$g_{2}^{2} / g_{3} \simeq 0.045$. Also,
$\epsilon_{b} \simeq 0.22$, therefore, $g_{2}^{2} / g_{3} \simeq 0.0413$.
We have used $g_{2}^{2} / g_{3} = 0.04$ in our calculations.

By fitting the experimental convective current at
$\epsilon=0.11$ (from Fig. 1 in \cite{re:bo91}), we obtain,
$(J_{conv})_{roll}  \simeq 0.16 = \frac{2\epsilon}{3g_{3}}$ or
$g_{3} \simeq 0.458$.
By fitting the experimental convective current at
$\epsilon=0.02$ (from Fig. 1 in \cite{re:bo91}) and by
using $g_{2}^{2}/g_{3}=0.045$, we obtain,
\begin{equation}
\label{eq-028}
(J_{conv})_{hex} \simeq 0.04 =
3\left[
(\sqrt{2}(g_{2}^{2}/\sqrt{g_{3}})^{1/2}+\sqrt{2g_{2}^{2}/g_{3}+30\epsilon})
/(15 \sqrt{g_{3}}) \right ] ^{2},
\end{equation}
or, $g_{3} \simeq 0.426$.

We next discuss the calculation of the numerical parameters
in the GSH equation related to large scale mean flow.
The dynamical equations are,
\begin{equation}
\label{eq-031}
\tau_{0}(\frac{\partial \psi}{\partial t}
 + \vec{U} \cdot \nabla \psi )=
\left[ \epsilon -  \frac{\xi_{0}^{2}}{4q_{c}^{2}} \left( \nabla^{2}
+ q_{c}^{2} \right)^{2} \right] \psi
-g_{2} \psi^{2} - g_{3} \psi^{3},
\end{equation}
and,
\begin{equation}
\left[ \frac{\partial }{\partial t} -Pr( \nabla^{2}-b^{2} ) \right]
\nabla^{2} \xi =
g_{m}\left[ \nabla(\nabla^{2} \psi) \times \nabla \psi \right]
\cdot \hat{e}_{z},
\end{equation}
where mean flow velocity $\vec{U}$,
\begin{equation}
\vec{U}=(\partial_{y} \xi) \hat{e}_{x} - (\partial_{x} \xi) \hat{e}_{y}.
\end{equation}
Here $\epsilon=\frac{R}{R_{c}}-1$ is the reduced Rayleigh number.
$R_{c}$ is the critical Rayleigh number for an infinite system,
and $Pr$ is the Prandlt number. The constants
$\tau_{0}$ and $\xi_{0}$ are the characteristic time and length scales,
$q_{c}$ is the critical wave number and $g_{2}$, $g_{3}$
are the nonlinear coupling constants. $b^{2}$ is an unknown constant.
Here $g_{m}=R_{c}<u_{0\perp}(z)^{2}>/(q_{c}^{2} <u_{0z}(z)\theta_{0}(z)>)$.
We have used the fact that near onset \cite{re:ah81},
\begin{equation}
\left[ V_{\perp}(\vec{r},z,t), V_{z}(\vec{r},z,t),
\theta(\vec{r},z,t) \right]
\approx \frac{1}{C}
\left[ u_{0\perp}(z)\partial_{\perp},u_{0z}(z),
\theta_{0}(z) \right] \psi(\vec{r},t)
\end{equation}
Here $\vec{V}=(V_{\perp}, V_{z})$ are the velocity field and $\theta$
is the deviation of the temperature from the linear conduction
profile.
The functions $u_{0\perp}(z), u_{0z}(z)$ and $\theta_{0}(z)$ are the
first eigenmodes in the vertical direction of the order parameter.
Here $\vec{r}$ denotes the two-dimensional horizontal coordinate.
The constant $C=\sqrt{<u_{0z}(z)\theta_{0}(z)>/R_{c}}$.
The symbol $< \ \ >$ means here an average over the vertical direction.
now rescale
$      \vec{r}^{ \prime } =q_{c} \vec{r},$
$       t^{\prime}=\frac{q_{c}^{2} \xi_{0}^{2}} {4 \tau_{0}} t ,$
$       \psi^{\prime}=\frac{2 \sqrt{g_{3}} }{q_{c} \xi_{0}} \psi ,$
$       \epsilon^{\prime}=\frac{4}{q_{c}^{2} \xi_{0}^{2}} \epsilon, $
$       \xi^{\prime}=\frac{g_{3}}{g_{m} \tau_{0} q_{c}^{2}} \xi$.
The rescaled GSH equation with large scale flow field becomes,
\begin{equation}
\label{eq-033}
\frac{\partial \psi^{\prime} }{\partial t^{\prime}}
 + g_{m}^{\prime} \vec{U}^{\prime} \cdot \nabla \psi^{\prime} =
\left[ \epsilon^{\prime} - \left( \nabla^{\prime ~ 2}
+ 1 \right)^{2} \right] \psi^{\prime}
-g_{2}^{\prime} \psi^{\prime ~ 2} -\psi^{\prime ~ 3},
\end{equation}
\begin{equation}
\left[ \frac{\partial }{\partial t^{\prime}}
-Pr^{\prime}( \nabla^{\prime ~ 2}-c^{\prime ~ 2} ) \right]
\nabla^{\prime ~ 2} \xi^{\prime} =
\left[ \nabla^{\prime}(\nabla^{\prime ~ 2} \psi^{\prime})
\times \nabla^{\prime} \psi^{\prime} \right]
\cdot \hat{e}_{z},
\end{equation}
where
$ g_{2}^{\prime}=(2 g_{2})/(\xi_{0}q_{c}\sqrt{g_{3}}),$
$ g^{\prime}_{m}=(4\tau_{0}^{2}g_{m}q_{c}^{2})/(g_{3}\xi_{0}^{2})$,
$ Pr^{\prime}=(4\tau_{0}/\xi_{0}^{2})Pr,$
and $ c^{\prime ~ 2}=b^{2}/q_{c}^{2}.$
 From the experiments on CO$_{2}$ \cite{re:bo91},
we use $ R_{c}=1707, q_{c}=3.117,
\xi_{0}^{2}=0.148, Pr=1,
\tau_{0}=0.07693
(\tau_{0}=\frac{Pr+0.5117}{19.65Pr}$   for
rigid-rigid boundary conditions),
$g_{m}=2.52$ (for rigid-rigid boundary conditions),
$g_{2}^{2}/g_{3}=0.04$, and $g_{3}=0.458$. We finally have
$ g_{2}^{\prime}=0.335,$ $ Pr^{\prime}=2.079,$
$ g^{\prime}_{m}=8.548,$
$ \epsilon^{\prime}=2.7818 \epsilon,$
and $c^{\prime ~ 2}$ is an unknown constant.

To recapitulate, we use in our numerical solution $g^{\prime}_{2}=0.35$,
$g^{\prime}_{m}=50$, $c^{\prime ~ 2}=10$ and $Pr^{\prime}=1.0$.
For these values the stability boundaries of the various patterns are,
\begin{equation}
\begin{array}{l}
\epsilon^{\prime}_{a}=-0.008167, ~ {\rm or}, ~ \epsilon_{a}=-0.00293,\\
\\
\epsilon^{\prime}_{r}=0.1633, ~ {\rm or}, ~ \epsilon_{r}=0.0587,\\
\\
\epsilon^{\prime}_{b}=0.6533, ~ {\rm or}, ~ \epsilon_{b}=0.2348,\\
\\
\epsilon^{\prime}_{T}=-0.00726, ~ {\rm or}, ~ \epsilon_{T}=-0.00261,\\
\\
\epsilon^{\prime}_{T^{\prime}}=0.24, ~ {\rm or}, ~
\epsilon_{T^{\prime}}=0.0863.\\
\end{array}
\end{equation}
Note that we have omitted the primes in the
Equations (\ref{eq:nonsh}-\ref{eq:bcs}).

\newpage

{\bf Figure captions}

Figure 1. Nucleation of a pattern of hexagonal symmetry
in a square cell of aspect ratio $\Gamma=128$. The values of the
parameters used are $g_{2} = 0.35, \epsilon = 0.01$ and $f=0$. In a
small square region at the center of the cell, $\epsilon = 0.055$. The
time shown is $t = 611$ and dark (white) areas represent regions in
which $\psi$ is positive (negative).

Figure 2. Hexagonal pattern obtained from a random initial condition
in a cylindrical cell of aspect ratio $\Gamma=64$. The values of
the parameters used are $g_{2} = 0.35$, $g_{m}=50$ and $\epsilon = 0.1$. A
forcing field localized at the boundary $f = 0.1$ has
been used. Four different times, (a),
$t=1.3$; (b), $t=51.2$; (c), $t=1049.7$; and, (d), $t=4649.7$ are shown.

Figure 3. This figure shows the early stages of hexagon to roll transition
produced by suddenly
changing $\epsilon$ from $\epsilon=0.1$ to $\epsilon=0.3$, in a cylindrical
cell of aspect ratio $\Gamma=64$. The initial condition is
the uniform hexagonal pattern shown in Fig. 2a.
Four different times,
(a), $t=480$; (b), $t=720$; (c), $t=840$; and, (d), $t=960$ are shown.
Rolls appear near defects and sidewall boundaries and
spread through the cell as the transition proceeds.

Figure 4. Formation of a rotating three-armed spiral
in a cylindrical cell of aspect ratio $\Gamma=64$, with $g_{2} = 0.35$,
$g_{m}=50$, $\epsilon = 0.3$ and $f=0.1$. The configurations shown are
continuation of those in Fig. 3. The times shown are
(a), t=1400; (b), t= 7440; (c), t=18000; (d), t=41040; (e), t=48240;
and, (f), t=64080. The final rotating spiral pattern is shown in (e) and
(f) is stable.

Figure 5. Formation of a hexagonal pattern from a three-armed
rotating spiral (Fig. 4(f)), with $g_{2}=0.35$, $g_{m}=50$, $\epsilon=0.1$
and f=0.1. An interesting feature to note is that the orientation
of the hexagonal domains tends to be normal to the rolls.
Four configurations during th early transition period are shown here:
(a), t=649; (b), t=1299; (c), t=2130; and, (d), t=3090.

Figure 6. Formation of a two-armed rotating spiral from random
initial conditions, with $g_{2}=0.35$, $g_{m}=50$, $\epsilon=0.3$ and
$f=0.1$. The configurations are
shown at (a), t=70.1; (b), t=190.7; (c), t=310.7; and, (d), t=1990.7.

Figure 7. Formation of a zero-armed,
and a one-armed rotating spiral pattern obtained from a random
initial condition with different values of $\epsilon$ from Fig. 6.
All the other parameters are the same as in Fig. 6.
(a), a zero-armed spiral with $\epsilon=0.22$,
and (b), a one-armed spiral with $\epsilon=0.26$.

Figure 8. Study of the stability of the spiral pattern.
We use the one-armed spiral shown in Fig. (7b) as
initial condition, switch $g_{2}$ from $g_{2}$=0.35 to $g_{2}$=0.0 and
keep all the other parameters unchanged.
Concentric rolls propagate inwards as the core of
the one-armed spiral starts shrinking. Finally the one-armed
spiral decays to a set of concentric rolls.
The times shown are (a), t=0; (b), t= 180; (c), t= 252; and,
(d), t= 288.

Figure 9. Convective heat current $J_{conv}$ versus the number
of arms for the same value of $\epsilon = 0.26$
The average convective current
decreases with increasing the number of arms in the spiral.


\begin{thebibliography}{99}
\bibitem{re:note92} We note that the vertical-diffusion time is
$\frac{t_{h}}{\Gamma^{2}}$.
\end{thebibliography}
\end{document}